\documentclass[aps,pre,twocolumn,superscriptaddress,showpacs,a4paper]{revtex4-1}
\usepackage{graphicx}
\usepackage{amssymb,amsmath}
\usepackage{wasysym}
\usepackage{textcomp}
\usepackage[dvips,usenames]{color}

\def\r{{\mathbf{r}}}
\def\ro{{\mathbf{r}_0}}

\begin{document}

\title{Geometry-induced fluctuations of olfactory searches in bounded domains}

\author{Juan Duque Rodr\'iguez}
\email{jrduque@ucm.es}
\affiliation{Laboratory of Physical Properties TAGRALIA,
Technical University of Madrid, 28040 Madrid, Spain}
\affiliation{CEI Campus Moncloa, UCM-UPM, Madrid, Spain.}

\author{David G\'omez-Ullate}
\email{dgomezu@ucm.es}
\affiliation{CEI Campus Moncloa, UCM-UPM, Madrid, Spain.}
\affiliation{Instituto de Ciencias Matem\'aticas (CSIC-UAM-UC3M-UCM),  C/ Nicolas Cabrera 15, 28049 Madrid, Spain.}  
\affiliation{Department of Theoretical Physics II, Complutense
  University of Madrid, 28040 Madrid, Spain}

\author{Carlos Mej\'{\i}a-Monasterio}
\email{carlos.mejia@upm.es}
\affiliation{Laboratory of Physical Properties TAGRALIA,
Technical University of Madrid, 28040 Madrid, Spain}
\affiliation{CEI Campus Moncloa, UCM-UPM, Madrid, Spain.}

\date{\today}

\begin{abstract}
  In olfactory  search an immobile target emits  chemical molecules at
  constant rate.  The molecules are transported by the medium which is
  assumed to be turbulent.  Considering a searcher able to detect such
  chemical signals and whose motion follows the infotaxis strategy, we
  study the  statistics of the  first-passage time to the  target when
  the searcher moves on  a finite two-dimensional lattice of different
  geometries.   Far  from  the  target,  where  the  concentration  of
  chemicals is low  the direction of the searcher's  first movement is
  determined by  the geometry  of the domain  and the topology  of the
  lattice, inducing strong fluctuations on the average search time
  with  respect to  the initial position  of the  searcher. The
  domain  is partitioned in  well defined  regions characterized  by the
  direction of the  first movement. If the  search starts over  the interface  between two different regions, large fluctuations in the search time are observed.
\end{abstract}


\maketitle

\section{Introduction}

Optimal olfactory  searches are essential for the  survival of insects
and animals in search for food and mating. These searches are based on
the capacity of these organisms to detect the chemical signals emitted
by  the target,  usually  at distances  that  are further  away to  be
visually located.  In recent years several olfactory search strategies
have  been proposed  to  model the  way  in which  such organisms  are
capable        of       finding       their        targets       \cite
{Russell,osh1,Revelli,ralfb,ol1a,gel,Vergassola}.          To
accomplish  a successful  search the  strategy has  to be  reliable to
reach  the  target  most  of  the times,  robust  against  errors  and
efficient  in   the  sense  of  exhibiting   reasonably  short  search
times. From a  mathematical point of view, an  optimal strategy is one
that minimizes the search time, failure probability or any appropriate
cost function.

The complexity of  an olfactory search is in  general a consequence of
the complexity  of the transport  of the chemical signals  through the
environment.   It is natural  to expect  that organisms  using optimal
strategies  will  have a  survival  advantage  with  respect to  other
competitors  that   do  not.   However,  biological   success  in  one
environment will be unsuccessful in  others as the optimal search will
depend on the environment and on how limited the information about the
properties of the transport is.  At low Reynolds numbers, molecular
diffusion  dominates, and  the concentration  of the  chemical signals
used in  the search are smooth  and more importantly,  static. In this
regime,  searches  up-gradient  in  the concentration  field  such  as
chemotaxis, are  successful.  This completely  exploitative strategy is
used  by bacteria  in search  for  food supplies,  phages to  localize
bacteria  and  euchariotic  cell  organisms  in  general.   At  larger
Reynolds numbers the  environment becomes turbulent, the concentration
field  breaks into  complex filamentous  structures  and concentration
decays  rapidly with  the distance  from the  source. In  this regime,
chemotaxis is ineffective since any local maxima, will attract the
searcher, keeping it from ever reaching the global maxima that marks
the position of the target.

In turbulent  environments the  searcher will need  to deal  with very
limited dynamically evolving information  to solve the task. Few years
ago, Vergassola, {\em et al}, proposed a strategy for olfactory search
called infotaxis \cite{Vergassola}.   They analyzed the search process
from an information theory point of view, as an inverse problem in
which the searcher needs to  decode partial and noisy messages to find
the position of  the emitter. As the searcher  moves, the
information  accumulated by the  detections of the  emitted chemicals is
used  to iteratively refine  the instantaneous  posterior distribution
function  for  the  probability  to  find the  target  at  a given position
$\mathbf{r}_0$, and to decide the further movements of the searcher as
those that maximize the local gain of information. Infotaxis has shown
to be  successful and robust,  leading to narrow distributions  for the
search time,  even if  the knowledge of  the environment is  not exact
\cite{Vergassola,Barbieri,Masson}.

In this paper we are concerned about the fluctuations of the search time
for infotaxis.  An optimal  search minimizing the fluctuations is more
reliable as then one would  expect that any search trajectory would be
completed in a time similar to the mean search time.  We will show for
infotaxis,  that when  the  search  process is  defined  on a  bounded
domain, the  fluctuations of  the search time  strongly depend  on the
geometry of  the domain and the  topology of the lattice  on which the
searcher evolves.  At the beginning  of the search, when no detections
have  yet  occurred,  one  would  expect that  the  searcher  dynamics
correspond to a random explorer.  However, here we show that the first
movements of the  searcher are not stochastic but  fully determined by
the geometry and the topology of the domain.  leading to a complicated
dependence of the search time  on the initial searcher's position. The
domain is partitioned in terms of the initial searcher's behaviour and
we find that the fluctuations  of the search time are largely enhanced
when  the search  starts  along the  interface  between partitions  of
different behaviour.

The   paper   is   organized   as   follows:   for   completeness,   in
section~\ref{sec:infotaxis}  we briefly  review  the infotaxis  search
algorithm and our implementation.  In  section~\ref{sec:inipos} the dependence  of the search
time on the initial position  of the searcher is analyzed, showing that the domain is decomposed in regions according to the first deterministic step. Furthermore,
in section~\ref{sec:geom}  we show that the boundary of these regions  depends  not only on the geometry of  the domain but also
on the topology of  the lattice and the parameters of the transport process. The geometry-induced fluctuations are studied in  section~\ref{sec:Pw} in terms  of the simultaneity  of the first   passage    process.    Our   findings    are   summarized   in
section~\ref{sec:concl}.

\section{Infotaxis search strategy}
\label{sec:infotaxis}

Infotaxis  is  a  Bayesian  olfactory  search strategy  in  which  the
information  released by the  target source  is transported  through a
turbulent   environment   \cite{Vergassola}.    An  infotactic   searcher
initially explores the space  collecting information, which is encoded
in  the  trace  of  detections  of molecules  emitted  by  the  source
$\mathcal{T}_t$.

This  information  is used  to  reconstruct,  at  each time  $t$,  the
posterior probability distribution $P_t(\mathbf{r}_0)$ for the unknown
position  of  the  source  $\mathbf{r}_0$.  Both  $\mathcal{T}_t$  and
$P_t(\mathbf{r}_0)$  are time-varying  quantities that  are constantly
updated.  The  posterior probability distribution, referred  to as the
\textit{belief   function},  depends   on  the   rate   of  detections
$R(\mathbf{r},\mathbf{r}_0)$  {\em  i.e.},   the  expected  number  of
detections at position $\mathbf{r}$, given that the emitting source is
located at  $\mathbf{r}_0$. This function $R(\mathbf{r},\mathbf{r}_0)$
incorporates our  modelling of the  natural transport process,  and it
can  be either  postulated  from rather  general  hypothesis, or  else
adapted dynamically as the searcher learns from the environment. As it is
customary, in this  paper we choose the first  option, and assume that
$R(\mathbf{r},\mathbf{r}_0)$ is the solution of an advection-diffusion
partial differential equation  \cite{Vergassola}, which in two spatial
dimensions takes the following explicit form:
\begin{equation}
R(\mathbf{r},\mathbf{r}_0)= \frac{\gamma}{\ln \left( \frac{\lambda}{a} \right)} e^{\frac{V\left( y_0 - y \right)}{2D}} K_0 \left( \frac{|\mathbf{r} - \mathbf{r}_0|}{\lambda} \right) \ ,
\end{equation}
where $\gamma$ is the rate of emission of volatile particles,  $D$ is their isotropic effective diffusivity, $a$ is the characteristic size of the detector, $K_0$ is the  modified Bessel function of order 0, and $\lambda$ the \textit{correlation length}, given by
\begin{equation}\label{eq:lambda}
 \lambda = \left( \frac{D \tau}{1 + \frac{V^2 \tau}{4D}} \right)^{1/2}
\end{equation}
where $\tau$ the finite lifetime of the emitted molecules, and $V$ the 
mean current or wind (which blows, without loss of generality, in the negative
$y$-direction). The correlation length $\lambda$ can be interpreted as the mean distance travelled by a volatile particle before it decays. 
In this paper we set the values of $a$ and $D$ to unity.

We stress  that no perfect  model of the environment  is realistically
available  in  concrete  applications   and  that  the  rate  function
$R(\mathbf{r},\mathbf{r}_0)$  will  only be  an  approximation to  the
actual transport  process. However,  the performance of  the infotaxis
strategy is still  acceptable despite an incorrect modelling of
the medium, which happens, for instance, if the parameters used by the
searcher in its model of  the environment differ from the real parameters
of the transport process. 


The simplest  model for the time correlation  of successive detections
is to assume that they  are independent. In a real turbulent transport
process  there   are  clearly  spatio-temporal   correlations  between
detections, as when a plume of odor particles hits the searcher, it is
more likely that in a short time after it will have a higher chance of
detection.  However,  ignoring this fact and  assuming no correlations
works surprisingly well for the Bayesian searcher to update its belief
function   of  the   source's  position   \cite{Vergassola}.  Assuming
independence of  the detections, the stochastic  process is Poissonian
and therefore  the probability distribution  at time $t$  posterior to
experiencing a trace $\mathcal{T}_t$ is given by 
\begin{equation}
 P_t(\ro) = \frac{\mathcal{L}_{\ro}(\mathcal{T})}{\int \mathcal{L}_{x}(\mathcal{T}) dx} \ ,
\end{equation}
where
$\mathcal{L}_{\ro} = e^{-\int_0^t R(\r(t')|\ro)dt' \prod_{i=1}^H  R(\r(t_i)|\ro)}$ 
and $H$ is  the total number of detections registered  by the searcher at
successive times   $(t_1,\dots,t_H)$.    Assuming   that   the   detections   are
uncorrelated, the evolution of the probability map can be written as a
Markovian update
\begin{equation}\label{prob-update}
 P_{t+\delta t}(\ro) = P_t(\ro) \frac{1}{Z_{t+\delta t}}e^{R(\r(t+\delta t)|\ro)\delta t}R^\eta(\r(t+\delta t)|\ro) \ ,
\end{equation}
where  $\eta$  is the  number  of  detections  during the  short  time
interval $\delta  t$ and $Z_{t+\delta t}$ is  a normalization constant
with respect to $\ro$. 
So far, this is a  purely Bayesian approach to infer the position
of the source from the trace of detections.  What is characteristic of
Infotaxis  is the  criterion for  the searcher's  motion:  rather than
moving towards  the most  probable source position,  at each  time the
searcher moves so that the expected information gain is maximized.

In  what follows  we assume  that the  searcher  moves  on  a regular
lattice with given topology.  Suppose the searcher is at position $\r$
at time $t$  with belief function $P_t(\ro)$. At  time ${t+\delta t} $
its new position $\r'$ can be any of the neighbouring lattice sites or
  $\r'=\r$ (no motion).   For  each  of  the  possible  motions  $\r'$
infotaxis  estimates the  Shannon  entropy associated  to the  updated
belief function  $P_{t+\delta t}(\ro)$ and  chooses from them  the one
for which the decrease in  entropy is highest.  The expected variation
of entropy upon moving to $\r'$ is
\begin{equation}\label{deltaS}
\overline{\Delta S}(\r\rightarrow\r') = -P_t(\r^\prime) S +
(1 - P_t(\r'))\left[\sum_{k=0}^\infty \rho_k(\r') \Delta S_k \right]
\end{equation}
where $\rho_k(\r')  = h(\r')^k  e^{-h(\r')}/k!$ is the  probability of
having $k$ detections during the time $\delta t$, where $h(\r') =
\delta  t \int  P_t(\ro) R(\r'|\ro)  d\ro$ is  the mean  number of
detections at position $\r'$.

Taking  $\eta=k$ in  \eqref{prob-update} allows  to  estimate the
updated  belief function  at  each  of the  sites  $\r'$ provided  $k$
detections were made  in the last time interval.   The Shannon entropy
of each of these probability densities is denoted by $S_k$ and $\Delta
S_k$ appearing in \eqref{deltaS} is the difference between the Shannon
entropies of the estimated $P_{t+\delta t}$ and the current $P_t$.

The first term on the right hand side of \eqref{deltaS} corresponds to
the event  of finding the source  at the neighboring  $\r'$ point. The
second term takes  into account the entropy decrease  if the source is
not  found  at $\r'$.  The  function to  maximize  at  each time  step
\eqref{deltaS} therefore  considers a balance  between exploration and
exploitation   of   the   acquired   information,  as   discussed   in
\cite{Vergassola}. The  algorithm evaluates \eqref{deltaS}  at each of
the possible $\r'$ and chooses  to move in the direction where $\Delta
S$ is highest, i.e. where  the largest expected decrease in entropy is
obtained.

We have numerically solved the infotaxis algorithm described above for
a searcher  moving on  a finite two-dimensional  lattice $\mathrm{L}$
with square, triangular and hexagonal topology. The lattice is bounded
by a  subset of lattice points $\delta\mathrm{L}$  that limit the
region in which the search process takes place. When  the searcher is
at  position $\r\in\mathrm{L}$  it can  move  to any  of the  nearest
neighbouring  points to  $\r$, unless  $\r\in\delta\mathrm{L}$,  in which
case the motion is allowed only over the subset of nearest
neighbouring points that are contained in $\mathrm{L}$. We set the
separation between lattice sites to one. Furthermore, we set the time
between two movements of the searcher to $\delta t =1$.

At  time  $t=0$,   the  searcher  starts  at  $\r_0\in\mathrm{L}$.
Initially  the belief function  is the  uniform distribution  over the
domain $\mathrm{L}$,  corresponding to  a complete lack  of knowledge
about  the position  of the  source and  therefore of  maximal entropy
$-\ln  N$,  with   $N$  the  total  number  of   points  contained  in
$\mathrm{L}$.  

During the  search process  the searcher moves,  registers detections,
and  updates its  belief  function according  to  the rules  described
above.  As detections occur, the belief function starts to concentrate
around the true  position of the source.  The search  ends at the time
$t$  at which  the Shannon  entropy reaches  a value  below  a certain
threshold, which we take to  be $10^{-4}$. Note that this
first hitting  time criterion for the  end of the  search is different
than the first passage criterion considered in \cite{Vergassola}.  For
small values of the emission  rate $\gamma \approx 1$, the search time
obtained  from both  criteria  is different  by  at most  one unit  of
time. However, at  high values of $\gamma$ the  first passage time may
diverge while our  first hitting time criterion is  still able to find
the  source. In  such situations  the search  ends and  the  source is
located even if the searcher  has not physically reached the source.
In the  rest of the paper we  study the statistics of  the search time
with respect  to the geometry of the  boundary $\delta\mathrm{L}$ and
the  topology of  the  lattice. In  this  paper we  consider only  low
emission rates and set $\gamma=2.4$. Therefore, our present results are
also valid for the search time corresponding to the first passage time.

\section{Dependence of the search time on the searcher's initial position}
\label{sec:inipos}

In this section  we study the dependence of the mean  search time as a
function  of  the initial  position  of  the  searcher.  We  consider  an
infotactic    searcher   in   a    square   lattice    of   dimensions
$100\times100$. The source is located  at the center of the lattice and
we consider first the case without advection ($V=0$).

\begin{figure}[!t]
  \centerline{
  \includegraphics*[width=0.4\textwidth]{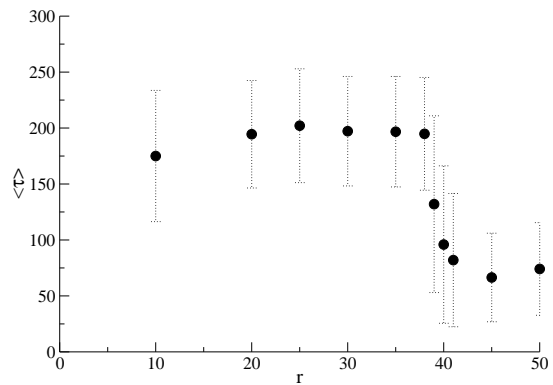}
  }
  \caption{Mean search time  $\langle\tau\rangle$ as a
    function of  the starting position  $(0,r)$ of the  searcher.  The
    source  is  located at  the  center  of  the lattice  $(0,0)$  and
    $\lambda=50$. The error bars corresponds to the standard deviation.}
  \label{fig:tau}
\end{figure}

It is natural to expect that  the search time increases monotonically with the initial  distance of the searcher to the source \cite{Vergassola,MassonPNAS}. This  is indeed the case  in an infinite
lattice. Here instead we first  show in Fig.~\ref{fig:tau} that due to
the  finite size of  the domain with reflecting
boundaries,  the search  time  exhibits a  complex  dependence with  the
initial  distance to  the  source.  Each  point in  Fig.~\ref{fig:tau}
corresponds  to the  mean search  time  $\langle\tau\rangle$, averaged
over $10^4$ trajectories starting at the same initial position $(0,r)$.

From Fig.~\ref{fig:tau} one observes two separate regions: when the
 starting position is $\lesssim 40$, the mean
search time is $\langle\tau\rangle\approx200$ and shows a small variation
with the distance $r$. Surprisingly, we find that when the searcher
starts further away at $r > 40$, the mean search time drops down to
$\langle\tau\rangle\approx75$. Furthermore, near the transition
distance $r\approx40$, the fluctuations of $\tau$ are larger. 

The  key observation  comes from  realizing  that, in  the absence  of
detections, the first step of an infotactic searcher is not random but
deterministic. With  an initial  uniform belief function  the searcher
moves in  the direction in which  the expected gain  of information is
maximal and, in the absence  of information about the source location,
the initial direction depends only  on the parameters of the transport
process and  the geometry of  the domain, but  not on the  position of
source.

\begin{figure}[!t]
  \centerline{\includegraphics*[width=0.195\textwidth]{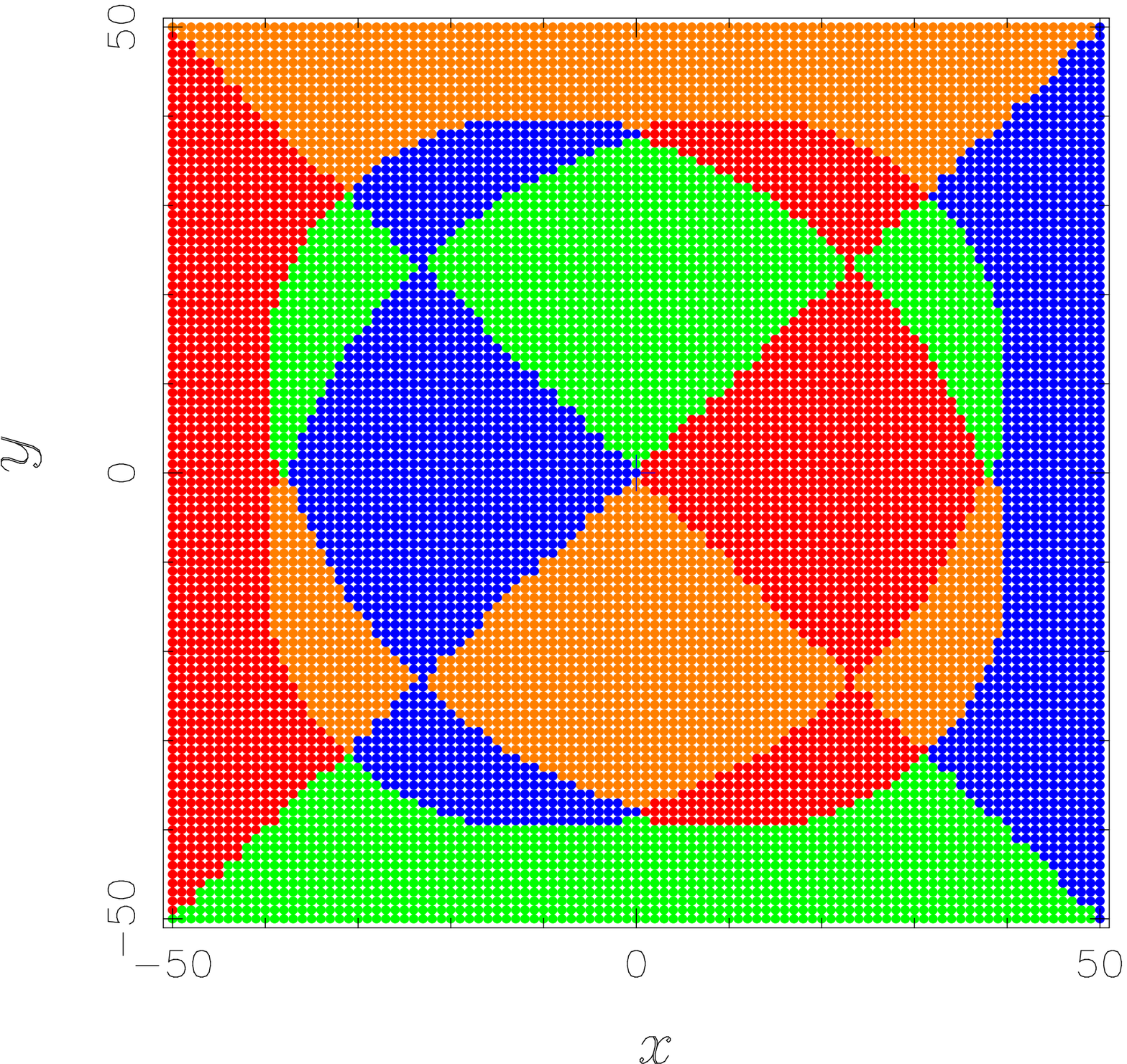}
 \includegraphics*[width=0.043\textwidth]{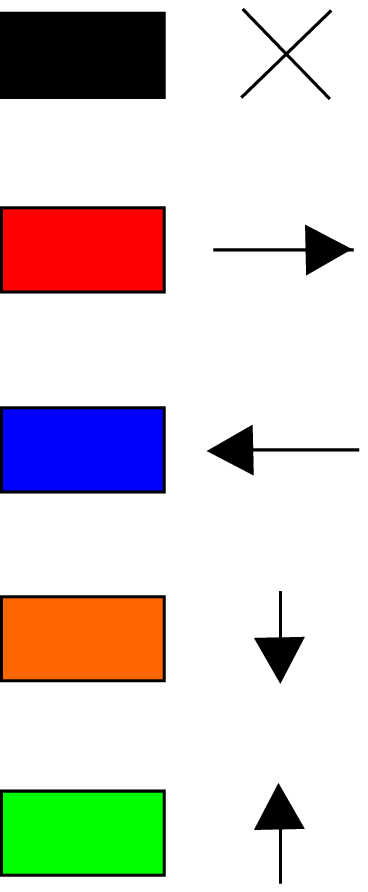}
  \includegraphics*[width=0.21\textwidth]{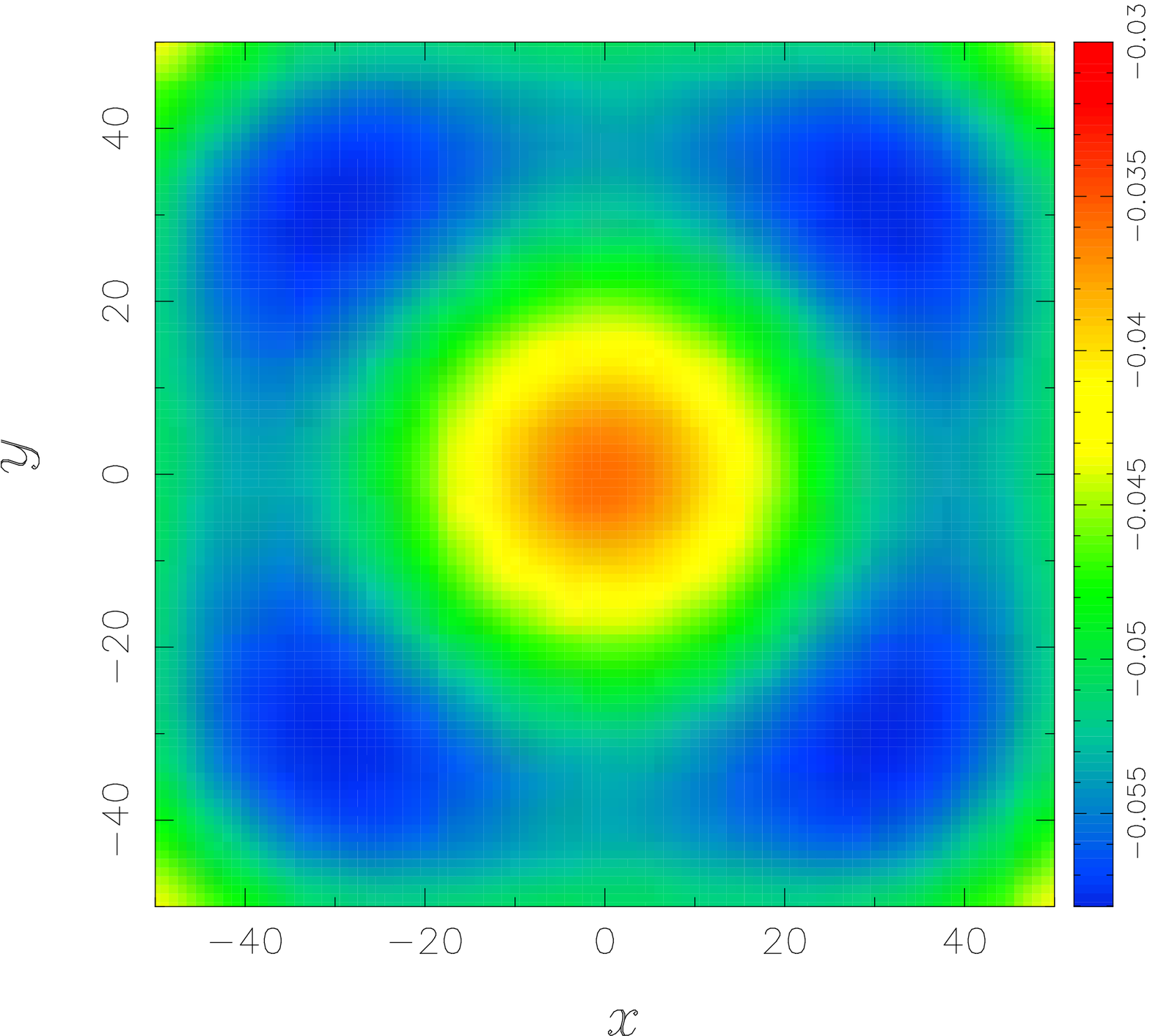}}
\caption{(Color online) Left panel:  First-step decision map for $V=0$
  and $\lambda=50$  in a  $100\times100$ square lattice.  Right panel:
  Expected entropy decrease $\Delta S$ after the first step.}
  \label{fig:nowind}
\end{figure}

At dilute conditions, {\em i.e.} far away from the source, it is clear
that the initial step will persist at longer times, until the searcher
starts detecting the molecules emitted by the source. As a consequence,
the direction of the first step will determine the region the searcher
explores first and thus, will have a considerable effect on the value
of $\langle\tau\rangle$. To understand the consequences of the first
step we show in Fig.~\ref{fig:nowind} (left panel),  a color code map
for the first step  of infotaxis. The color of each lattice site
represents the direction of the first step of the searcher when its
initial position is at the same lattice site. It is noticeable that although standing still is always an option for the infotaxis algorithm, we find that this option is never optimal for the first step.

From   Fig.~\ref{fig:nowind} it is   evident  that the first step in Infotaxis is deterministic and the direction is strongly influenced  by  the  geometry  of the  lattice boundary.   The complex  dependence of  $\langle\tau\rangle$  with the initial distance to  the source of  Fig.~\ref{fig:tau} can now be explained from the structure of the first-step decision map.  At
short distances $r \lesssim 40$  the first step is upwards (away  from the  source).  On  the contrary,  at larger
distances  $r\gtrsim 40$  the first  step is  downwards  (towards the  source). Depending  on the  direction of  its first
step,  the  searcher will  explore  different  regions  of the  search
domain. If the first step is away from the source, it is less probable for the searcher to have detections and it will keep moving until it reaches the vicinity of the boundary.  This first  exploration depends on the  size  and
geometry of the domain and the time invested on
it will be added  to the total search time. Instead, when the searcher
starts close to the boundary, it moves towards the center of the
lattice, and since the source is placed there, the searcher will start
detecting much sooner than in the previous case and the mean search
time is smaller. 

From the previous explanation, it is important to remark that the dependence
$\langle\tau\rangle$ on the initial distance $r$ shown in Fig. \ref{fig:tau} will be different if the source is not located at the center, but the structure of the  first-step decision map shown in Fig. \ref{fig:nowind} does not depend on the position of the source.

\begin{figure}[!t]
  \centerline{\includegraphics*[width=0.45\textwidth]{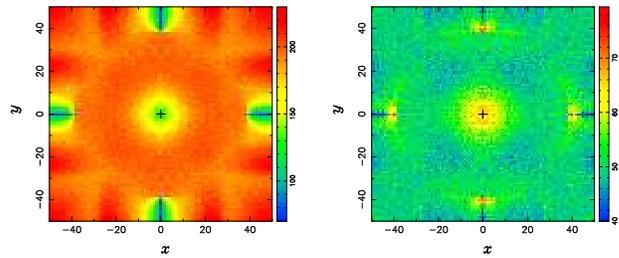}}
  \caption{(Color online) Mean  search time $\langle\tau\rangle$ (left
    panel) and its  standard deviation (right panel) as  a function of
    the  initial position,  averaged over  $400$ trajectories  and the
    same parameters as in Fig.~\ref{fig:nowind}.}
  \label{fig:nowind_}
\end{figure}

To understand  the different regions in Fig.~\ref{fig:nowind} we need to look at the expression for
the  expected entropy  change.  Since  the initial  belief map  is the
uniform   distribution,  only   the   sum  inside   the  brackets   in
Eq.~\eqref{deltaS}  changes  for  different  choices of  $\r'$.   This
quantity depends  on the position  through the functions  $h(\r')$ and
$R(\r'|\r_0)$. In the right panel of Fig.~\ref{fig:nowind} we show for
each lattice site,  the maximal expected decrease of  entropy as given
by Eq.~\eqref{deltaS}.  There are four  minima on each diagonal of the
square domain which  act as attractors for the  infotaxis, and the the
first step of the searcher is determined by them.  The locus of points
for which  the maximal  decrease in entropy  occurs for  two different
directions define  the smooth curves separating  the different regions
in the left  panel of Fig.~\ref{fig:nowind}.  The symmetry  of the map
reflects  the  symmetry  of the  boundary,  and  in  this case  of  no
advection ($V=0$) the isotropy of the transport process.  
Therefore, the first-step decision map is determined by  the geometry of
the  domain via the way in which the expected maximal entropy
decrease. This generates a well defined partition in the search domain
into regions with a structure that inherit, in the absence of wind,
the same symmetry as that of the boundary.

With  the   source  located   at  the  center   of  the   lattice,  in
Fig.~\ref{fig:nowind_}  we  show, for  each  initial  position in  the
lattice, the  mean search  time $\langle\tau\rangle$ (left  panel) and
its  standard deviation  (right panel)  as a  color density  plot. The
regions  near the  midpoints  of the  edges  on the  boundary lead  to the
shortest  search  times.   As  discussed  above,  in  the  absence  of
detections, the initial first  steps direct the searcher trajectory to
the center  of the lattice,  and therefore, to  the region in  which a
larger number of detections is  expected. More interesting, it is clear
from the right  panel that the fluctuations of  the search time $\tau$
are  larger  over the  partition  curves  of  the first-step  decision
map.  The  fluctuations  will   be  discussed  in  greater  detail  in
section~\ref{sec:Pw}.

\section{Dependence on the geometry and topology of the domain}
\label{sec:geom}


In this  section we complete the study of the effects that different
boundary geometries, lattice topologies and the parameters of
the transport process have on the  structure of the first-step decision map,
and consequently on the average search time.

\begin{figure}[!t] 
  \centerline{\includegraphics*[width=0.45\textwidth]{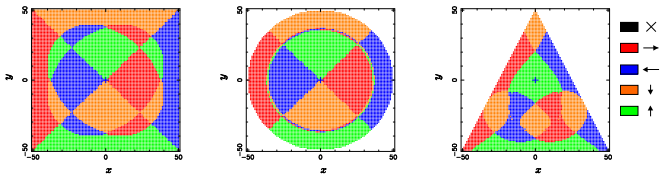} \vphantom{aaaa}\includegraphics*[width=0.029\textwidth]{Codigo_color_Square_Lattice.ps}}
  \caption{(Color   online)  First-step   decision   map  for   $V=0$,
    $\lambda=50$ and different boundary  geometries. The motion of the
    searcher is on a square lattice.  }
  \label{fig:boundary}
\end{figure}

We start by computing the  first-step decision map on a square lattice
with square,  circular and triangular boundary. The  results are shown
in  Fig.\ref{fig:boundary} for a  correlation length  $\lambda=50$. We
see how the partition of the search domain is affected by the geometry
and symmetry  of the boundary.  The conclusions drawn in  the previous
section apply  to all  geometries: in the  absence of  detections, the
first step of an infotactic  searcher is determined by the geometry of
the boundary.  The center of  the lattice acts as a repeller, i.e. all
searches starting at neighbouring points will have their first step away from the center. The boundary by construction is also repelling, but we see in Fig. \ref{fig:boundary} that there are also attracting points: four in the case of the square boundary, three for the triangular and none for the circular case. 

In terms of the expected entropy variation, this  means that  imposing reflecting  boundaries, an infotactic search will  always  produce maximal  entropy decrease  in regions that are  away from the center and the  boundary of the search domain (see the right panel of Fig. \ref{fig:nowind}).

\begin{figure}[!b]
  \centerline{ \includegraphics*[width=0.45\textwidth]{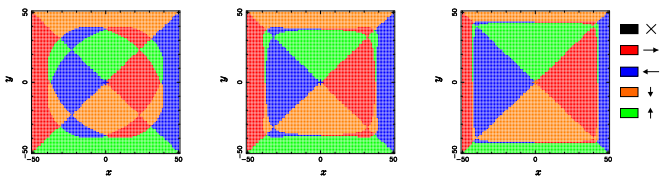}
  \includegraphics*[width=0.029  \textwidth]{Codigo_color_Square_Lattice.ps} }
\caption{(Color online)  First-step decision map for a square lattice with $V=0$, and three
  different  values  of the  correlation  length $\lambda=50$  (left),
  $\lambda=5$ (middle) and $\lambda=2$ (right). }
  \label{fig:lambda}
\end{figure}
 
In the  previous figures, the correlation length  $\lambda=50$ has the
same size as  the entire search domain. Roughly  speaking $\lambda$ is
the average  distance travelled by  the molecules emitted by  the source
before  decaying.   Therefore,   the  statistical  inference  done  by
infotaxis has an  effect over an area with  extension $\lambda$ around
the searcher.  In Fig.~\ref{fig:lambda} we study the dependence of the
first-step  decision  map  on  the  correlation length  for  a  square
lattice.  To  consider different values  of the correlation  length we
have  kept  fixed  the  effective  diffusivity $D=1$  and  varied  the
molecules' life  time $\nu$. As  the correlation length  decreases, the
central area of the decision map extends towards the boundary. Clearly,
the relevant parameter is the ratio between $\lambda$ and the size of
the lattice. 

The results  in the right panel of  Fig.~\ref{fig:lambda} suggest that
in  the limit  in  which the  lattice  size is  much  larger than  the
correlation length,  the central repeller dominates the  dynamics. As
expected, this happens irrespectively of the geometry of the boundary. The effect of the boundary becomes relevant only at distance $\lambda$ from it.
 Moreover, assuming  that the  source is  located  near the
center of the  lattice, the searcher
will start moving away form the source into regions where the detections become more rare, and this  outward motion  will only stop when the searcher
reaches the neighborhood of the boundary.  In particular, this means that
in an infinite domain infotactic searches in the absence of wind will
rarely be successful.

\begin{figure}[!t]
  \centerline{\includegraphics*[width=0.45\textwidth]{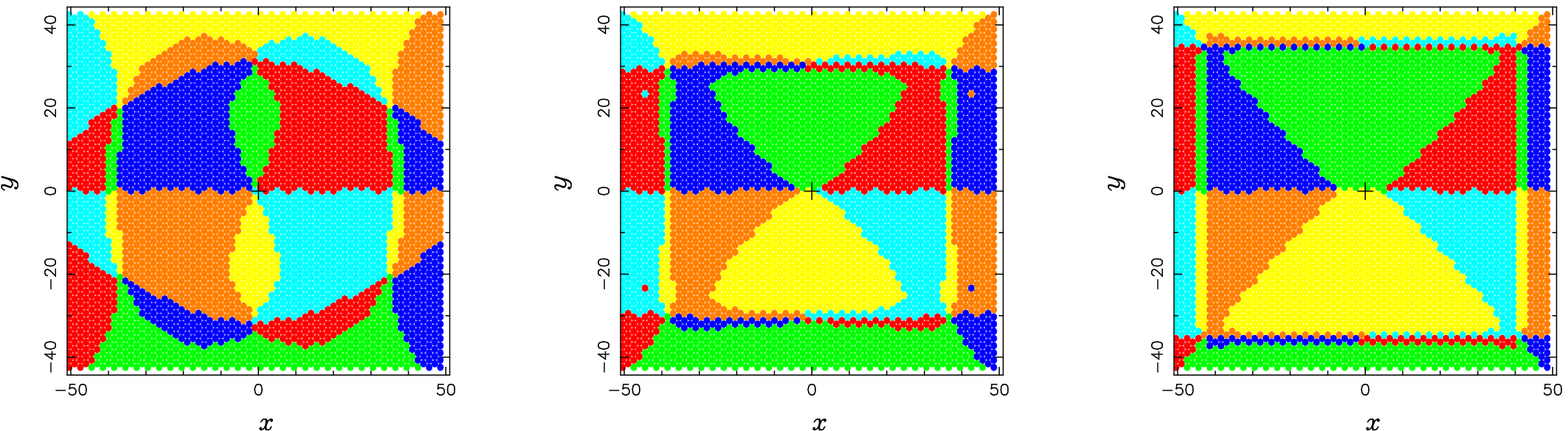} \vphantom{aa}\includegraphics*[width=0.026\textwidth]{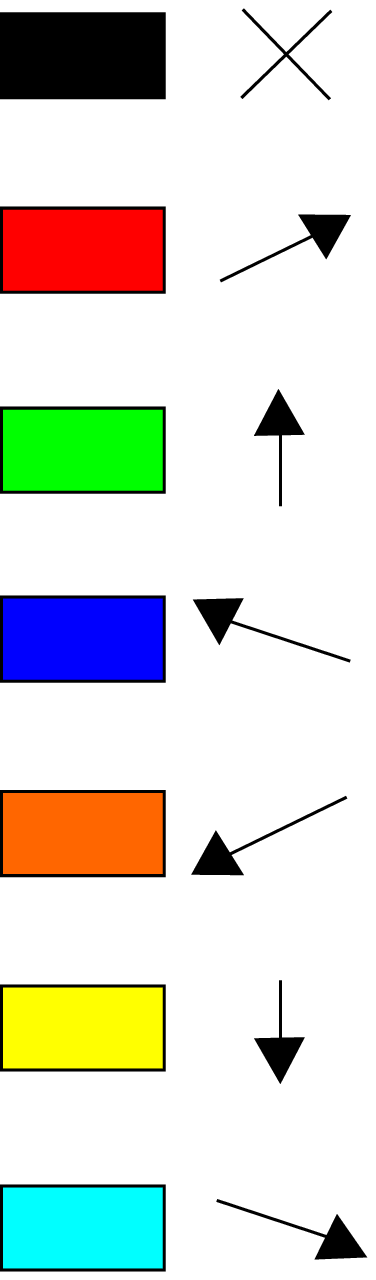}} 
    \caption{(Color online) First-step decision map for an hexagonal lattice with $V=0$,
    and three  different values of the  correlation length $\lambda=50$
    (left),  $\lambda=5$ (middle)  and  $\lambda=2$ (right).}
  \label{fig:lambda_hex}
\end{figure}

Besides the effect of boundary discussed above, the topology of the
lattice also influences the structure of the decision map. We illustrate
this in Fig.~\ref{fig:lambda_hex} where  the first-step decision map is
shown for an hexagonal lattice with a square boundary. The symmetry of
the partitions inherit the symmetry of the lattice topology, while the
rest of the properties found for the square lattice remain the same.

\begin{figure}[!t]
\centerline{\includegraphics*[width=0.15\textwidth]{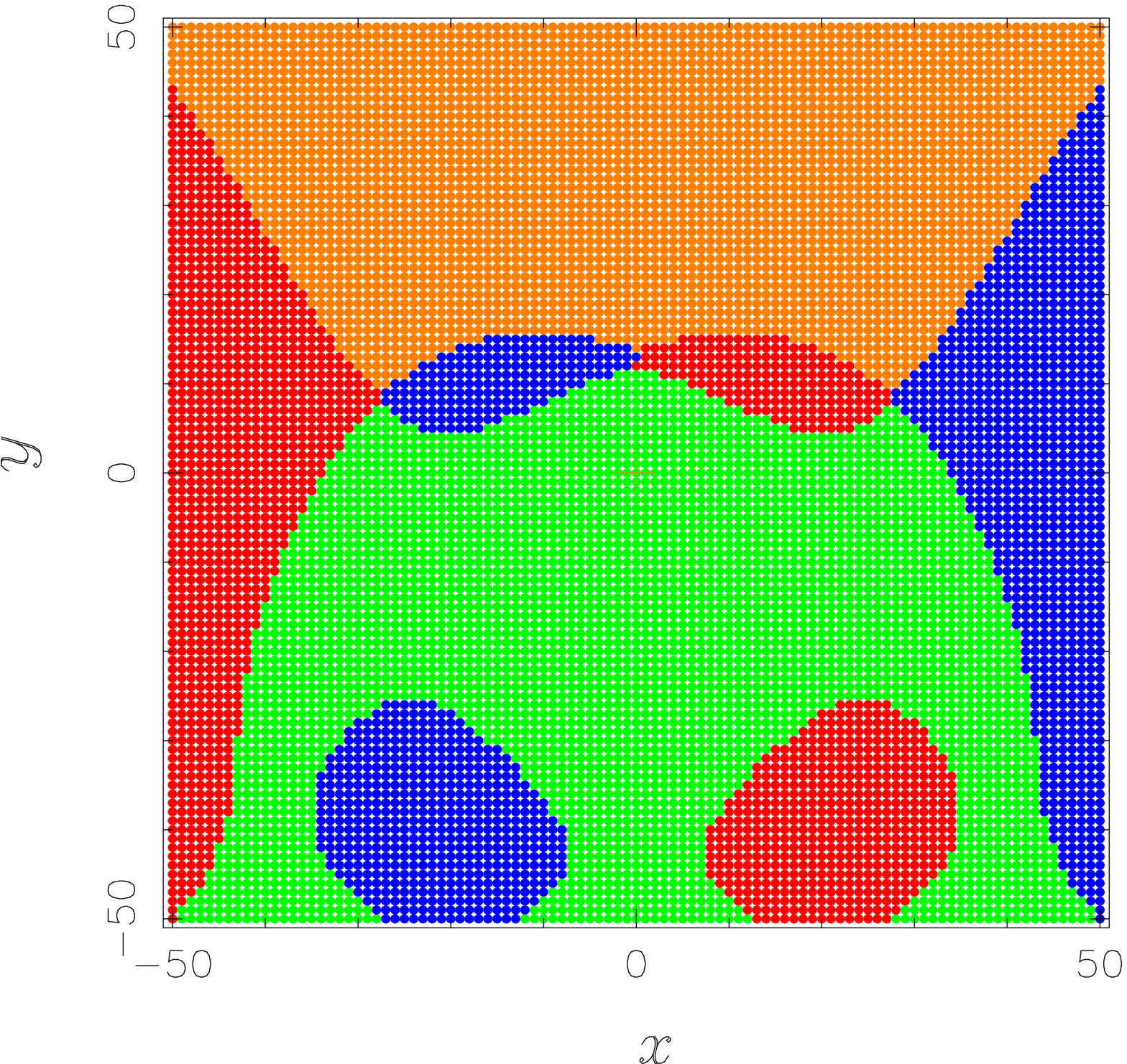}\includegraphics*[width=0.15\textwidth]{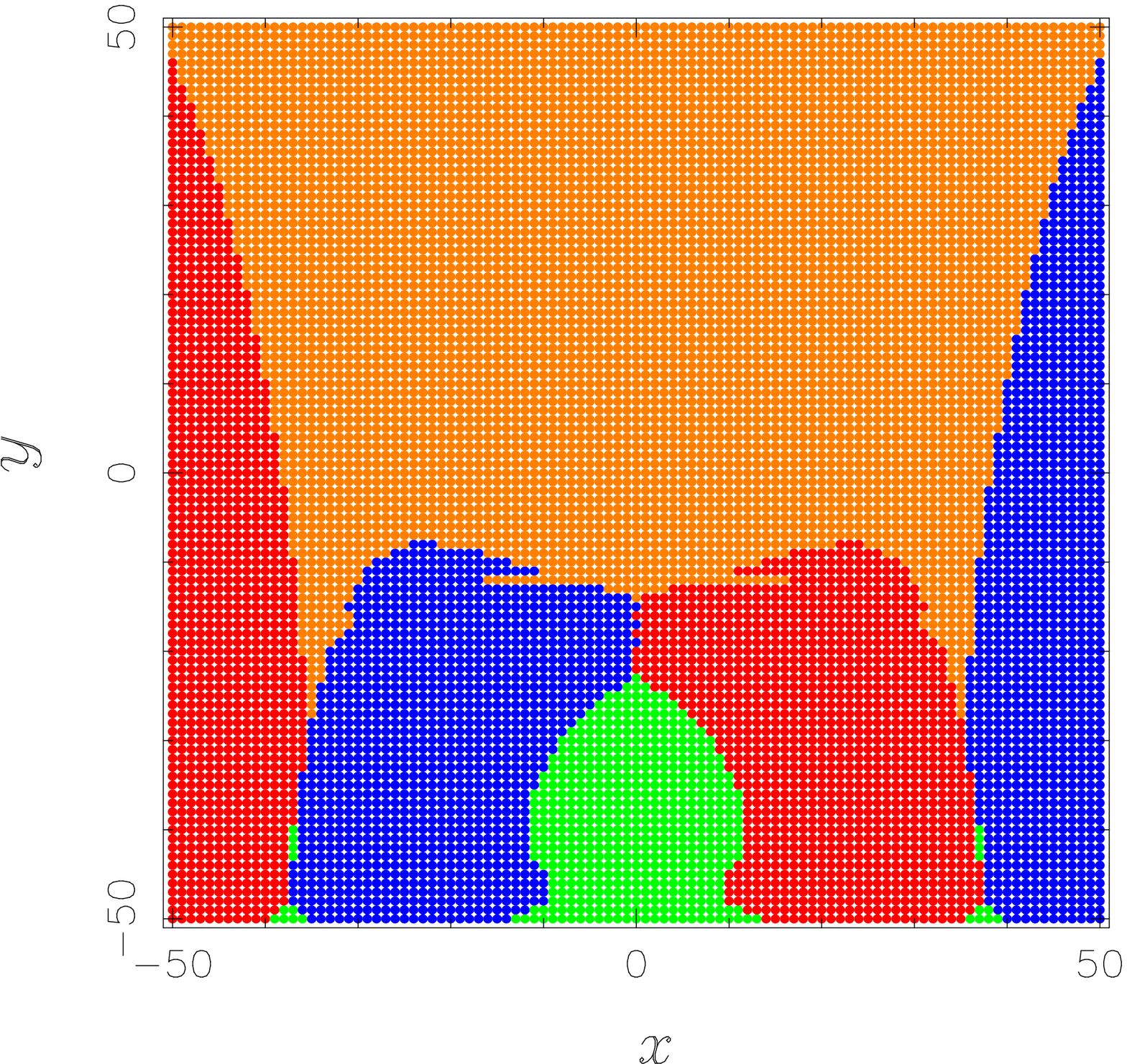}\includegraphics*[width=0.15\textwidth]{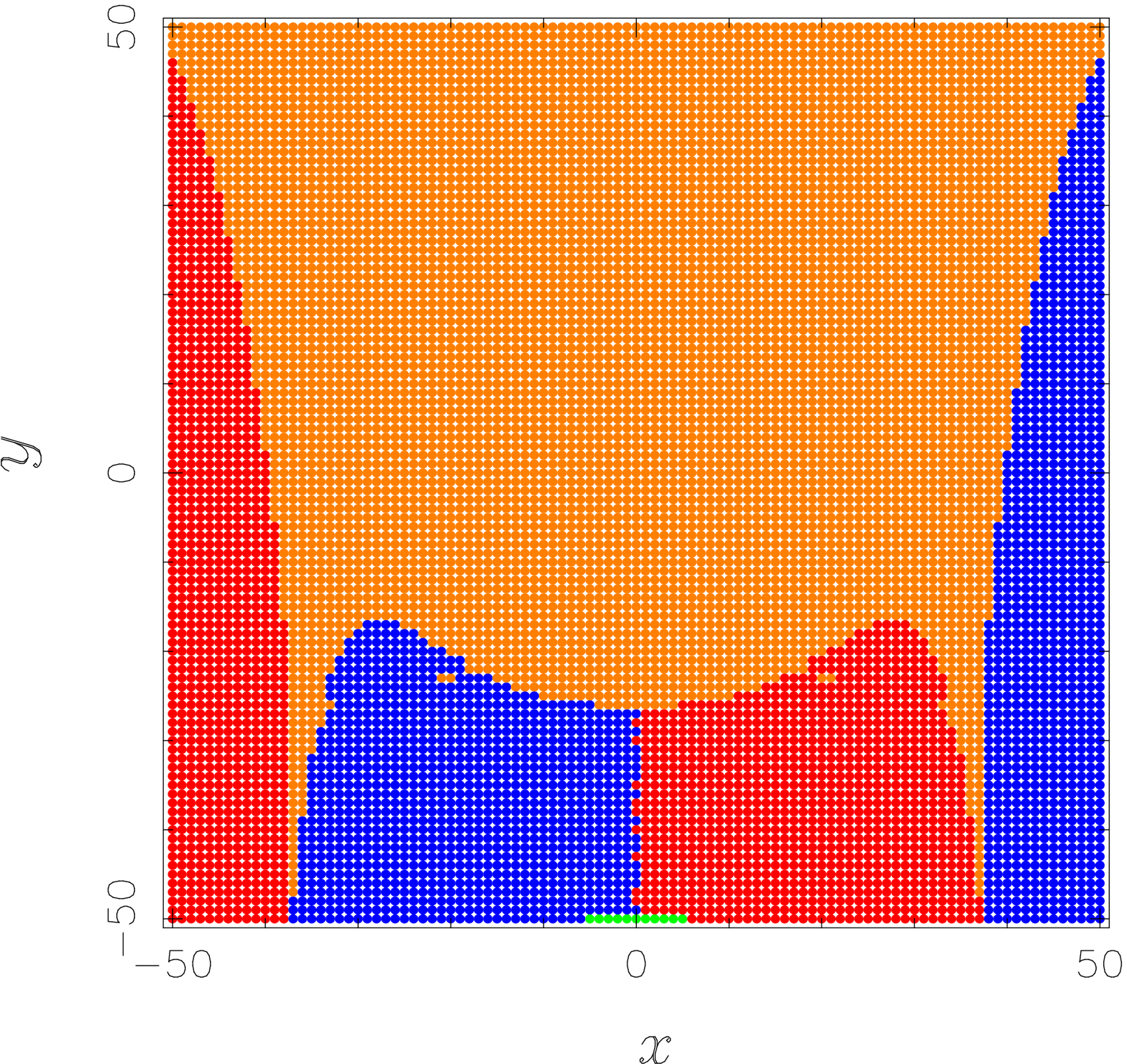} \vphantom{aaaa}\includegraphics*[width=0.033\textwidth]{Codigo_color_Square_Lattice.ps}}
\caption{(Color online)  First-step decision map for a square lattice
  and different values of the wind velocity: $V=0.1$ (left), $V=0.5$
  (center) and $V=0.9$ (right). The corresponding values of the
  correlation length determined by equation \eqref{eq:lambda} are $\lambda= 18.5$ (left), $\lambda=3.9$ (center) and $\lambda= 2.2$ (right).
  \label{fig:wind}}
\end{figure}

To conclude this section we consider the structure of the decision map
for the  case in which  an advection field  exists.  This is  shown in
Fig.~\ref{fig:wind} for  three different  values of the  wind velocity
$V$.   The presence of wind destroys the isotropy of the  problem.  A
non-zero $V$  yields additional  information on  the region  of the
search domain where the source may be located. Roughly speaking, at
large  P\'eclet numbers,  when the advection dominates  the molecular
diffusion, if the searcher does  not record detections then the source
is  most likely  located downwards  (recall that the wind  is directed
along the negative $y$-axis).   Conversely, if detections  occur then
the searcher will automatically  move upwards as the detected molecule
is expected to come from  that region.  This extra information clearly
appears in the decision map.  In the absence of detections, the region
in which the first step is downwards dominates at larger values
of $V$. The region corresponding to upwards motion shrinks to
zero and the regions in which the first move is left or right
remain important only near the boundary.

\section{Trajectory-to-trajectory fluctuations}
\label{sec:Pw}

In the last section we focus on the fluctuations of the search time in
terms  of the  statistics of  the \textit{simultaneity}  index of the
search as discussed in \cite{carlos,thiago}. 

Consider  two   independent  searchers  that  start  simultaneously  at  the  same
position ${\r_0}$, with corresponding search
times $\tau_1$ and $\tau_2$. One defines then the random variable
\begin{equation}
\label{def:omega}
\omega\equiv\frac{\tau_1}{\tau_1+\tau_2},
\end{equation}
such   that   $\omega$   ranges   in  the   interval   $[0,1]$.    The
\textit{simultaneity      index\/}       $\omega$      measures      the
\textit{likelihood\/}  that both  searches  end simultaneously, and
thus it sheds light about the size of the fluctuations of the search
time and the robustness of the search in general.

When $\omega\approx1/2$ any typical  search will end at similar times,
meaning that the  most probable value for the search  time is close to
its  statistical  mean.  On  the  contrary,  when $\omega\approx0$  or
$\omega\approx1$ it means that the  search will end at very dissimilar
times, so  that large  fluctuations in the  search time  are expected.
The statistics of the simultaneity  index is a quantitative measure of
the  robustness  of  a  search  process when  finding  the  source  at
reasonably similar times is important. These statistics have been used
in  the  past in  the  analysis  of  random probabilities  induced  by
normalisation  of self-similar  L\'evy processes~\cite{iddo1},  of the
fractal  characterisation of Paretian  Poisson processes~\cite{iddo2},
and of the  so-called Matchmaking paradox~\cite{iddo3,iddo4}, and more
recently,  to quantify  sample-to-sample fluctuations  in mathematical
finances~\cite{samor2,hol},    chaotic    systems~\cite{schehr},   the
analysis  of distributions  of the  diffusion coefficient  of proteins
diffusing  along DNAs\cite{boyer1,boyer1a}  and optimal  estimators of
the     diffusion     coefficient     of     a     single     Brownian
trajectory.\cite{boyer2}.

In section~\ref{sec:inipos}, we have observed that the fluctuations of
the search time depend on the position from which the search
starts. We showed in the right panel of Fig.~\ref{fig:nowind_} that
when the search starts over a line separating different regions of the
decision map the standard deviation is largest. In this section, we
explore this issue using the statistics of the simultaneity index.

\begin{figure}[!t]
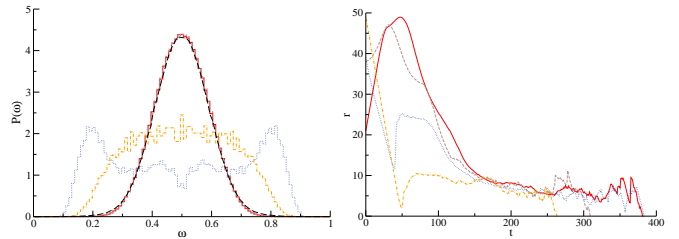

  \centerline{\includegraphics*[width=0.24\textwidth]{Pw.eps}
\includegraphics*[width=0.24\textwidth]{r.eps}}
  \caption{(Color  online)  Similarity  index  statistics (left panel) and average distance from the source as a
    function of  time (right panel),  for four
    different  initial positions $(0,r)$:  $r=20$ (red  solid), $r=38$
    (brown   dashed),  $r=40$   (blue  dotted)   and   $r=50$  (orange
    dotted-dashed).  These results were obtained for a square lattice,
    with $V=0$ and $\lambda=50$ averaged over $10^4$  trajectories.
  \label{fig:Pw}
  }
 
\end{figure}

We have numerically  computed the probability distribution $P(\omega)$
for different initial positions of the searcher and the source located
at  the   origin.   The  histograms   shown  in  the  left   panel  of
Fig.~\ref{fig:Pw}  correspond   to  some  of   the  initial  positions
considered in  Fig.~\ref{fig:tau}. For $r < 40$,  $P(\omega)$ is found
to be  a unimodal,  bell-shaped function with  a maximum at  $\omega =
1/2$.  This  means that  two different infotactic  searchers beginning
their trajectory within  this region will most likely  employ the same
time to find the source for  the first time. For $r > 40$, $P(\omega)$
remains  unimodal, but  with a  larger dispersion  than  before.  More
interestingly, for $r = 40$, when  the initial position lays on one of
the separatrix  curves, $P(\omega)$ exhibits a  bimodal, M-shaped form
with a local minimum at $\omega  = 1/2$ and two maxima at around $0.2$
and $0.8$1,  meaning that most  likely, different searchers  will find
the  source at  quite  different  times, which  in  turn implies  that
trajectory-to-trajectory fluctuations are  expected to be large.  This
is   exactly   what   it    has   been   qualitatively   observed   in
Fig.~\ref{fig:nowind_}

One can grasp the origin of these large fluctuations by looking at how
the    searcher   approaches    to   the    source   in    time.    In
the right panel of Fig.~\ref{fig:Pw}  we show how the distance from the source evolves
on   average.   In  agreement with  the  first-step decision  map
discussed  in  section~\ref{sec:inipos},  the searchers  with  initial
position near the center of  the lattice move outwards from the center
(and  from the  source)  for a  time  $\approx 50$,  after which  they
approach the source monotonically  until detection become frequent. In
the opposite case in which the searcher's initial position is far from
the center, they rapidly approach the source. 

As a final remark, we have computed the distribution
$P(\omega)$ averaged over all initial positions in the lattice (see the
black dashed curve in the left panel of Fig.~\ref{fig:Pw}). The bell-shaped distribution
indicates that infotaxis is a robust search algorithm in bounded
domains, even when the exact initial position of the search is not accessible.

\section{Conclusions}

We have studied  the infotaxis search strategy in  finite domains with
a reflecting boundary. We  find that the time  at which infotactic
searchers locate the source shows a complex dependence on the starting position.
We explain this dependence by noting that in the absence of detections the first step of infotaxis is deterministic, and we study the decision map for this first step. The search domain is partitioned into regions on which the first step is constant, and the boundaries of these regions are determined by the geometry of the boundary, the topology of the lattice and in general by the parameters of the transport process.
When the starting position falls near a separatrix curve, the fluctuations around the mean search time increase. These higher fluctuations can also be explained in terms of the initial step.

When the correlation length is much smaller than the domain size, the effectiveness of infotaxis will decrease. When the correlation length is comparable to the domain size, infotaxis remains on average a robust search strategy.

\label{sec:concl}


\acknowledgments 
This work has been supported by Grant n. 245986 of the EU-project Robots Fleets for Highly Agriculture and Forestry Management. 
JDR was also supported by a PICATA predoctoral fellowship of the Moncloa Campus of International Excellence (UCM-UPM). The research of DGU has been
supported in part by Spanish MINECO-FEDER Grants  MTM2012-31714 and FIS2012-38949-C03-01. We acknowledge the use of the UPC Applied Math cluster system for research computing (see http://www.ma1.upc.edu/eixam/index.html).

\end{document}